\documentclass[11pt,twoside]{article}

\usepackage{asp2014}

\aspSuppressVolSlug
\resetcounters

\begin{document}

\title{Montage and Radio Astronomy}


\author{G.~Bruce~Berriman,$^1$ John~C.~Good,$^2$ Ian~Heywood,$^3$ and R.~Moseley $^2$}
\affil{$^1$Caltech/IPAC-NExScI,~Pasadena,~CA~~91125,~USA;}~~~~~~~\email{gbb@ipac.caltech.edu}

\affil{$^2$Caltech/IPAC-NExScI,~Pasadena,~CA~~91125,~USA}

\affil{$^3$Department~of~Physics,~University~of~Oxford,~Keble~Road,~Oxford,~OX1~3RH,~UK}


\paperauthor{G.~B.~Berriman}{gbb@ipac.caltech.edu}{0000-0001-8388-534X}{Caltech}{IPAC/NExScI}{Pasadena}{CA}{91125}{USA}
\paperauthor{J.~C.~Good}{jcg@ipac.caltech.edu}{0009-0003-3906-719X}{Caltech}{IPAC/NExScI}{Pasadena}{CA}{91125}{USA}
\paperauthor{I.~Heywood}{}{ian.heywood@physics.ox.ac.uk}{0000-0001-6864-5057}{University~of~Oxford}{Department~of~Physics}{Oxford}{OX1~3RH}{UK}
\paperauthor{R.~Moseley}{moseley@ipac.caltech.edu}{0000-0002-0759-0475}{Caltech}{IPAC/NExScI}{Pasadena}{CA}{91125}{USA}



\begin{abstract}

The Montage Image Mosaic Engine, first released in 2002, has found applicability across the electromagnetic spectrum to support data processing and visualization. This broad applicability has come about through its design as an Open Source ANSI-C toolkit (and Python binary extensions), with independent components to perform each step in the creation of a mosaic and with support for all WCS extensions. This design enables easy integration into custom environments, workflows and pipelines, and is the principal reason for its long lifetime. Here we emphasize the growing use of Montage in radio astronomy (37 peer-reviewed papers since 2020), and will focus on three high-profile applications: 
(1) Analysis of observations made with SKA precursor experiments, such as MeerKAT and the Murchison Wide-field Array,
(2) Faraday tomography of LOFAR Two-Metre Sky Survey data (LoTSS-DR2), which explores the structure of the local interstellar medium, and (3) Identification of fast radio bursts.

\end{abstract}



\section{Take-up in Radio Astronomy}

For 22 years, the Open Source Montage Mosaic Engine \footnote{\url{https://github.com/Caltech-IPAC/Montage}} has found applicability across the electromagnetic spectrum to create mosaics and visualizations (\citet{2017PASP..129e8006B}. This take-up is a consequence of its design as a toolkit written in ANSI-C for performance and portability, with Python binary extensions of each toolkit module. 

In the past four years, Montage has found particular applicability in radio astronomy and in follow-up studies of radio observations, with 37 peer-reviewed papers reported since 2020, as of this writing \footnote{\url{https://zenodo.org/records/14226692}}. This paper describes three examples of science investigations in radio astronomy and one example of applicability in science outreach.

\section{Example 1: SKA Precursor Projects}
The SKA commissioned precursor projects to demonstrate its design and inform the development of its processing environment. The latter includes the development of the Containerized Automated Radio Astronomy Calibration (CARACal) pipeline as a general-purpose pipeline for radio interferometry data reduction \citep{2020ascl.soft06014J}, which emphasizes the re-use of existing packages. Montage has been incorporated into the MosaicQueen package, which is used to create 2D and 3D mosaics, as well as to perform primary beam correction for 2D and 3D images\footnote{\url{https://github.com/caracal-pipeline/MosaicQueen}}. Investigations taking advantage of Montage and CARACal include the SARAO MeerKAT 1.3 GHz Galactic Plane Survey \citep{2024MNRAS.531..649G}, HI Galaxy Signatures in the MeerKAT Galactic Plane Survey \citep{2024MNRAS.528..542K}, \citep{2024MNRAS.531.3486R}, and a study of ram pressure stripping in the Hydra I cluster \citep{2022A&A...668A.184H}.

Moreover, the MIGHTEE survey coadded MeerKAT pointings to create image mosaics in its studies of extragalactic deep fields, to study the formation and evolution of galaxies \citep{2024MNRAS.534...76H}. 

\section{Example 2: Faraday Tomography of LOFAR Two-Metre Sky Survey data (LoTSS-DR2)}
Montage was used to create a mosaic Faraday cube of LoTSS-DR2 data by applying a rotation measure synthesis algorithm, and subsequently construct Faraday moment maps to probe the magneto-ionized structure of our Galactic neighborhood (\citet{2024A&A...688A.200E} and references therein).

\section{Example 3:  Identification of Fast Radio Bursts}
Montage was used to coadd glare-subtracted follow-up images to host galaxies of FRBs found by the VLA and localized by ASKAP \citep{2022AJ....163...69B}.

\section{Science Outreach}
The IAU General Assembly in Cape Town, South Africa featured a hands-on satellite workshop entitled "Virtual Observatory Tools for Students and Educators in Africa." It emphasized all-sky science and the fusion of data from different archives and telescopes. Montage contributed a freely available Jupyter Notebook \footnote{\url  {https://github.com/Caltech-IPAC/MontageNotebooks/blob/main/MeerKAT.ipynb}} that creates a blend of a 3-color 2MASS mosaic of the Galactic Center with the MeerKAT 1.2 GHz mosaic, itself created with Montage \citep{2022ApJ...925..165H}. This image is shown in Figure 1.

\articlefigure{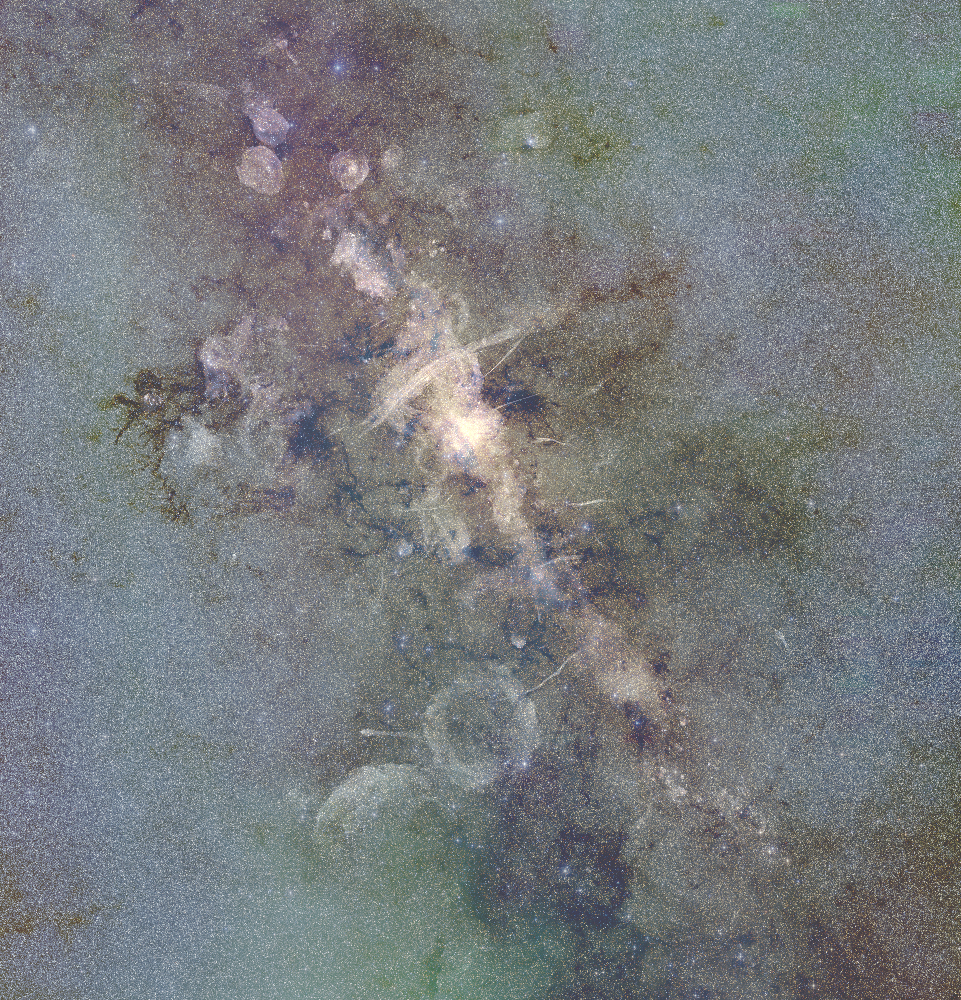}{ex_fig1}{A three-color blend of a three-color 2MASS mosaic of the Galactic Center with a 1.2 GHz mosaic of the same region acquired with MeerKAT.}

\acknowledgements
 This research made use of Montage. It is funded by the National Science Foundation under Grant Numbers ACI-1440620, 1642453, 1835379.

\bibliography{P302}  

\begin{thebibliography}{}
\expandafter\ifx\csname natexlab\endcsname\relax\def\natexlab#1{#1}\fi
\expandafter\ifx\csname url\endcsname\relax
  \def\url#1{\texttt{#1}}\fi
\expandafter\ifx\csname urlprefix\endcsname\relax\def\urlprefix{URL }\fi
\providecommand{\eprint}[2][]{\url{#2}}

\bibitem[{{Berriman} \& {Good}(2017)}]{2017PASP..129e8006B}
{Berriman}, G.~B., \& {Good}, J.~C. 2017, \pasp, 129, 058006. \eprint{1702.02593}

\bibitem[{{Bhandari} et~al.(2022){Bhandari}, {Heintz}, {Aggarwal}, {Marnoch}, {Day}, {Sydnor}, {Burke-Spolaor}, {Law}, {Xavier Prochaska}, {Tejos}, {Bannister}, {Butler}, {Deller}, {Ekers}, {Flynn}, {Fong}, {James}, {Lazio}, {Luo}, {Mahony}, {Ryder}, {Sadler}, {Shannon}, {Han}, {Lee}, \& {Zhang}}]{2022AJ....163...69B}
{Bhandari}, S., {Heintz}, K.~E., {Aggarwal}, K., {Marnoch}, L., {Day}, C.~K., {Sydnor}, J., {Burke-Spolaor}, S., {Law}, C.~J., {Xavier Prochaska}, J., {Tejos}, N., {Bannister}, K.~W., {Butler}, B.~J., {Deller}, A.~T., {Ekers}, R.~D., {Flynn}, C., {Fong}, W.-f., {James}, C.~W., {Lazio}, T. J.~W., {Luo}, R., {Mahony}, E.~K., {Ryder}, S.~D., {Sadler}, E.~M., {Shannon}, R.~M., {Han}, J., {Lee}, K., \& {Zhang}, B. 2022, \aj, 163, 69. \eprint{2108.01282}

\bibitem[{{Erceg} et~al.(2024){Erceg}, {Jeli{\'c}}, {Haverkorn}, {Gajovi{\'c}}, {Hardcastle}, {Shimwell}, \& {Tasse}}]{2024A&A...688A.200E}
{Erceg}, A., {Jeli{\'c}}, V., {Haverkorn}, M., {Gajovi{\'c}}, L., {Hardcastle}, M., {Shimwell}, T.~W., \& {Tasse}, C. 2024, \aap, 688, A200. \eprint{2406.14679}

\bibitem[{{Goedhart} et~al.(2024){Goedhart}, {Cotton}, {Camilo}, {Thompson}, {Umana}, {Bietenholz}, {Woudt}, {Anderson}, {Bordiu}, {Buckley}, {Buemi}, {Bufano}, {Cavallaro}, {Chen}, {Chibueze}, {Egbo}, {Frank}, {Hoare}, {Ingallinera}, {Irabor}, {Kraan-Korteweg}, {Kurapati}, {Leto}, {Loru}, {Mutale}, {Obonyo}, {Plavin}, {Rajohnson}, {Rigby}, {Riggi}, {Seidu}, {Serra}, {Smart}, {Stappers}, {Steyn}, {Surnis}, {Trigilio}, {Williams}, {Abbott}, {Adam}, {Asad}, {Baloyi}, {Bauermeister}, {Bennet}, {Bester}, {Botha}, {Brederode}, {Buchner}, {Burger}, {Cheetham}, {Cloete}, {de Villiers}, {de Villiers}, {du Toit}, {Esterhuyse}, {Fanaroff}, {Fourie}, {Gamatham}, {Gatsi}, {Geyer}, {Gouws}, {Gumede}, {Heywood}, {Hokwana}, {Hoosen}, {Horn}, {Horrell}, {Hugo}, {Isaacson}, {J{\'o}zsa}, {Jonas}, {Jordaan}, {Joubert}, {Julie}, {Kapp}, {Kriek}, {Kriel}, {Krishnan}, {Kusel}, {Legodi}, {Lehmensiek}, {Lord}, {Macfarlane}, {Magnus}, {Magozore}, {Main}, {Malan}, {Manley}, {Marais}, {Maree}, {Martens}, {Maruping}, {McAlpine},
  {Merry}, {Mgodeli}, {Millenaar}, {Mokone}, {Monama}, {New}, {Ngcebetsha}, {Ngoasheng}, {Nicolson}, {Ockards}, {Oozeer}, {Passmoor}, {Patel}, {Peens-Hough}, {Perkins}, {Ramaila}, {Ratcliffe}, {Renil}, {Richter}, {Salie}, {Sambu}, {Schollar}, {Schwardt}, {Schwartz}, {Serylak}, {Siebrits}, {Sirothia}, {Slabber}, {Smirnov}, {Tiplady}, {van Balla}, {van der Byl}, {Van Tonder}, {Venter}, {Venter}, {Welz}, \& {Williams}}]{2024MNRAS.531..649G}
{Goedhart}, S., {Cotton}, W.~D., {Camilo}, F., {Thompson}, M.~A., {Umana}, G., {Bietenholz}, M., {Woudt}, P.~A., {Anderson}, L.~D., {Bordiu}, C., {Buckley}, D.~A.~H., {Buemi}, C.~S., {Bufano}, F., {Cavallaro}, F., {Chen}, H., {Chibueze}, J.~O., {Egbo}, D., {Frank}, B.~S., {Hoare}, M.~G., {Ingallinera}, A., {Irabor}, T., {Kraan-Korteweg}, R.~C., {Kurapati}, S., {Leto}, P., {Loru}, S., {Mutale}, M., {Obonyo}, W.~O., {Plavin}, A., {Rajohnson}, S.~H.~A., {Rigby}, A., {Riggi}, S., {Seidu}, M., {Serra}, P., {Smart}, B.~M., {Stappers}, B.~W., {Steyn}, N., {Surnis}, M., {Trigilio}, C., {Williams}, G.~M., {Abbott}, T.~D., {Adam}, R.~M., {Asad}, K.~M.~B., {Baloyi}, T., {Bauermeister}, E.~F., {Bennet}, T.~G.~H., {Bester}, H., {Botha}, A.~G., {Brederode}, L.~R.~S., {Buchner}, S., {Burger}, J.~P., {Cheetham}, T., {Cloete}, K., {de Villiers}, M.~S., {de Villiers}, D.~I.~L., {du Toit}, L.~J., {Esterhuyse}, S.~W.~P., {Fanaroff}, B.~L., {Fourie}, D.~J., {Gamatham}, R.~R.~G., {Gatsi}, T.~G., {Geyer}, M., {Gouws}, M., {Gumede},
  S.~C., {Heywood}, I., {Hokwana}, A., {Hoosen}, S.~W., {Horn}, D.~M., {Horrell}, L.~M.~G., {Hugo}, B.~V., {Isaacson}, A.~I., {J{\'o}zsa}, G.~I.~G., {Jonas}, J.~L., {Jordaan}, J.~D.~B.~L., {Joubert}, A.~F., {Julie}, R.~P.~M., {Kapp}, F.~B., {Kriek}, N., {Kriel}, H., {Krishnan}, V.~K., {Kusel}, T.~W., {Legodi}, L.~S., {Lehmensiek}, R., {Lord}, R.~T., {Macfarlane}, P.~S., {Magnus}, L.~G., {Magozore}, C., {Main}, J.~P.~L., {Malan}, J.~A., {Manley}, J.~R., {Marais}, S.~J., {Maree}, M.~D.~J., {Martens}, A., {Maruping}, P., {McAlpine}, K., {Merry}, B.~C., {Mgodeli}, M., {Millenaar}, R.~P., {Mokone}, O.~J., {Monama}, T.~E., {New}, W.~S., {Ngcebetsha}, B., {Ngoasheng}, K.~J., {Nicolson}, G.~D., {Ockards}, M.~T., {Oozeer}, N., {Passmoor}, S.~S., {Patel}, A.~A., {Peens-Hough}, A., {Perkins}, S.~J., {Ramaila}, A.~J.~T., {Ratcliffe}, S.~M., {Renil}, R., {Richter}, L.~L., {Salie}, S., {Sambu}, N., {Schollar}, C.~T.~G., {Schwardt}, L.~C., {Schwartz}, R.~L., {Serylak}, M., {Siebrits}, R., {Sirothia}, S.~K., {Slabber},
  M.~J., {Smirnov}, O.~M., {Tiplady}, A.~J., {van Balla}, T.~J., {van der Byl}, A., {Van Tonder}, V., {Venter}, A.~J., {Venter}, M., {Welz}, M.~G., \& {Williams}, L.~P. 2024, \mnras, 531, 649. \eprint{2312.07275}

\bibitem[{{Hess} et~al.(2022){Hess}, {Kotulla}, {Chen}, {Carignan}, {Gallagher}, {Jarrett}, \& {Kraan-Korteweg}}]{2022A&A...668A.184H}
{Hess}, K.~M., {Kotulla}, R., {Chen}, H., {Carignan}, C., {Gallagher}, J.~S., {Jarrett}, T.~H., \& {Kraan-Korteweg}, R.~C. 2022, \aap, 668, A184. \eprint{2209.05605}

\bibitem[{{Heywood} et~al.(2024){Heywood}, {Ponomareva}, {Maddox}, {Jarvis}, {Frank}, {Adams}, {Baes}, {Bianchetti}, {Collier}, {Deane}, {Glowacki}, {Jung}, {Pan}, {Rajohnson}, {Rodighiero}, {Ruffa}, {Santos}, {Sinigaglia}, \& {Vaccari}}]{2024MNRAS.534...76H}
{Heywood}, I., {Ponomareva}, A.~A., {Maddox}, N., {Jarvis}, M.~J., {Frank}, B.~S., {Adams}, E.~A.~K., {Baes}, M., {Bianchetti}, A., {Collier}, J.~D., {Deane}, R.~P., {Glowacki}, M., {Jung}, S.~L., {Pan}, H., {Rajohnson}, S.~H.~A., {Rodighiero}, G., {Ruffa}, I., {Santos}, M.~G., {Sinigaglia}, F., \& {Vaccari}, M. 2024, \mnras, 534, 76. \eprint{2409.17713}

\bibitem[{{Heywood} et~al.(2022){Heywood}, {Rammala}, {Camilo}, {Cotton}, {Yusef-Zadeh}, {Abbott}, {Adam}, {Adams}, {Aldera}, {Asad}, {Bauermeister}, {Bennett}, {Bester}, {Bode}, {Botha}, {Botha}, {Brederode}, {Buchner}, {Burger}, {Cheetham}, {de Villiers}, {Dikgale-Mahlakoana}, {du Toit}, {Esterhuyse}, {Fanaroff}, {February}, {Fourie}, {Frank}, {Gamatham}, {Geyer}, {Goedhart}, {Gouws}, {Gumede}, {Hlakola}, {Hokwana}, {Hoosen}, {Horrell}, {Hugo}, {Isaacson}, {J{\'o}zsa}, {Jonas}, {Joubert}, {Julie}, {Kapp}, {Kenyon}, {Kotz{\'e}}, {Kriek}, {Kriel}, {Krishnan}, {Lehmensiek}, {Liebenberg}, {Lord}, {Lunsky}, {Madisa}, {Magnus}, {Mahgoub}, {Makhaba}, {Makhathini}, {Malan}, {Manley}, {Marais}, {Martens}, {Mauch}, {Merry}, {Millenaar}, {Mnyandu}, {Mokone}, {Monama}, {Mphego}, {New}, {Ngcebetsha}, {Ngoasheng}, {Ockards}, {Oozeer}, {Otto}, {Passmoor}, {Patel}, {Peens-Hough}, {Perkins}, {Ramaila}, {Ramanujam}, {Ramudzuli}, {Ratcliffe}, {Robyntjies}, {Salie}, {Sambu}, {Schollar}, {Schwardt}, {Schwartz}, {Serylak},
  {Siebrits}, {Sirothia}, {Slabber}, {Smirnov}, {Sofeya}, {Taljaard}, {Tasse}, {Tiplady}, {Toruvanda}, {Twum}, {van Balla}, {van der Byl}, {van der Merwe}, {Van Tonder}, {Van Wyk}, {Venter}, {Venter}, {Wallace}, {Welz}, {Williams}, \& {Xaia}}]{2022ApJ...925..165H}
{Heywood}, I., {Rammala}, I., {Camilo}, F., {Cotton}, W.~D., {Yusef-Zadeh}, F., {Abbott}, T.~D., {Adam}, R.~M., {Adams}, G., {Aldera}, M.~A., {Asad}, K.~M.~B., {Bauermeister}, E.~F., {Bennett}, T.~G.~H., {Bester}, H.~L., {Bode}, W.~A., {Botha}, D.~H., {Botha}, A.~G., {Brederode}, L.~R.~S., {Buchner}, S., {Burger}, J.~P., {Cheetham}, T., {de Villiers}, D.~I.~L., {Dikgale-Mahlakoana}, M.~A., {du Toit}, L.~J., {Esterhuyse}, S.~W.~P., {Fanaroff}, B.~L., {February}, S., {Fourie}, D.~J., {Frank}, B.~S., {Gamatham}, R.~R.~G., {Geyer}, M., {Goedhart}, S., {Gouws}, M., {Gumede}, S.~C., {Hlakola}, M.~J., {Hokwana}, A., {Hoosen}, S.~W., {Horrell}, J.~M.~G., {Hugo}, B., {Isaacson}, A.~I., {J{\'o}zsa}, G.~I.~G., {Jonas}, J.~L., {Joubert}, A.~F., {Julie}, R.~P.~M., {Kapp}, F.~B., {Kenyon}, J.~S., {Kotz{\'e}}, P.~P.~A., {Kriek}, N., {Kriel}, H., {Krishnan}, V.~K., {Lehmensiek}, R., {Liebenberg}, D., {Lord}, R.~T., {Lunsky}, B.~M., {Madisa}, K., {Magnus}, L.~G., {Mahgoub}, O., {Makhaba}, A., {Makhathini}, S., {Malan}, J.~A.,
  {Manley}, J.~R., {Marais}, S.~J., {Martens}, A., {Mauch}, T., {Merry}, B.~C., {Millenaar}, R.~P., {Mnyandu}, N., {Mokone}, O.~J., {Monama}, T.~E., {Mphego}, M.~C., {New}, W.~S., {Ngcebetsha}, B., {Ngoasheng}, K.~J., {Ockards}, M.~T., {Oozeer}, N., {Otto}, A.~J., {Passmoor}, S.~S., {Patel}, A.~A., {Peens-Hough}, A., {Perkins}, S.~J., {Ramaila}, A.~J.~T., {Ramanujam}, N.~M.~R., {Ramudzuli}, Z.~R., {Ratcliffe}, S.~M., {Robyntjies}, A., {Salie}, S., {Sambu}, N., {Schollar}, C.~T.~G., {Schwardt}, L.~C., {Schwartz}, R.~L., {Serylak}, M., {Siebrits}, R., {Sirothia}, S.~K., {Slabber}, M., {Smirnov}, O.~M., {Sofeya}, L., {Taljaard}, B., {Tasse}, C., {Tiplady}, A.~J., {Toruvanda}, O., {Twum}, S.~N., {van Balla}, T.~J., {van der Byl}, A., {van der Merwe}, C., {Van Tonder}, V., {Van Wyk}, R., {Venter}, A.~J., {Venter}, M., {Wallace}, B.~H., {Welz}, M.~G., {Williams}, L.~P., \& {Xaia}, B. 2022, \apj, 925, 165. \eprint{2201.10541}

\bibitem[{{J{\'o}zsa} et~al.(2020){J{\'o}zsa}, {White}, {Thorat}, {Smirnov}, {Serra}, {Ramatsoku}, {Ramaila}, {Perkins}, {Moln{\'a}r}, {Makhathini}, {Maccagni}, {Kleiner}, {Kamphuis}, {Hugo}, {de Blok}, \& {Andati}}]{2020ascl.soft06014J}
{J{\'o}zsa}, G. I.~G., {White}, S.~V., {Thorat}, K., {Smirnov}, O.~M., {Serra}, P., {Ramatsoku}, M., {Ramaila}, A. J.~T., {Perkins}, S.~J., {Moln{\'a}r}, D.~C., {Makhathini}, S., {Maccagni}, F.~M., {Kleiner}, D., {Kamphuis}, P., {Hugo}, B.~V., {de Blok}, W.~J.~G., \& {Andati}, L. A.~L. 2020, {CARACal: Containerized Automated Radio Astronomy Calibration pipeline}, Astrophysics Source Code Library, record ascl:2006.014

\bibitem[{{Kurapati} et~al.(2024){Kurapati}, {Kraan-Korteweg}, {Pisano}, {Chen}, {Rajohnson}, {Steyn}, {Frank}, {Serra}, {Goedhart}, \& {Camilo}}]{2024MNRAS.528..542K}
{Kurapati}, S., {Kraan-Korteweg}, R.~C., {Pisano}, D.~J., {Chen}, H., {Rajohnson}, S. H.~A., {Steyn}, N., {Frank}, B., {Serra}, P., {Goedhart}, S., \& {Camilo}, F. 2024, \mnras, 528, 542. \eprint{2312.05237}

\bibitem[{{Rajohnson} et~al.(2024){Rajohnson}, {Kraan-Korteweg}, {Chen}, {Frank}, {Steyn}, {Kurapati}, {Pisano}, {Staveley-Smith}, {Serra}, {Goedhart}, \& {Camilo}}]{2024MNRAS.531.3486R}
{Rajohnson}, S. H.~A., {Kraan-Korteweg}, R.~C., {Chen}, H., {Frank}, B.~S., {Steyn}, N., {Kurapati}, S., {Pisano}, D.~J., {Staveley-Smith}, L., {Serra}, P., {Goedhart}, S., \& {Camilo}, F. 2024, \mnras, 531, 3486. \eprint{2405.15629}

\end{thebibliography}


\end{document}